\IfFileExists{IEEEtran.cls}
  {\documentclass[conference,letterpaper]{IEEEtran}}
  {\documentclass[10pt,twocolumn,letterpaper]{article}
   \usepackage[margin=0.7in,columnsep=0.25in]{geometry}}
\makeatletter
\@ifclassloaded{IEEEtran}{}{%
  \newcommand{\IEEEauthorblockN}[1]{#1\\}
  \newcommand{\IEEEauthorblockA}[1]{{\small\itshape\begin{tabular}[t]{@{}c@{}}#1\end{tabular}}}
  \newenvironment{IEEEkeywords}{\par\noindent\textbf{\textit{Index Terms}}---}{\par\medskip}
  \renewcommand{\abstract}{\par\noindent\textbf{\textit{Abstract}}---\bfseries}
  
  \usepackage{titlesec}
  \renewcommand\thesection{\Roman{section}}
  \renewcommand\thesubsection{\Alph{subsection}}
  \titleformat{\section}{\centering\normalsize\scshape}{\thesection.}{0.5em}{}
  \titleformat{\subsection}{\normalsize\itshape}{\thesubsection.}{0.5em}{}
  \titlespacing*{\section}{0pt}{1.2ex}{0.8ex}
  \titlespacing*{\subsection}{0pt}{1.0ex}{0.5ex}
}
\makeatother

\usepackage[T1]{fontenc}
\usepackage{mathptmx}
\usepackage{amsmath,amssymb,amsfonts}
\usepackage{graphicx}
\usepackage{booktabs,multirow}
\usepackage{cite}
\usepackage{algorithm}
\usepackage{algpseudocode}
\usepackage{tikz}
\usetikzlibrary{arrows.meta,positioning,shapes.geometric,calc,fit}
\usepackage{xcolor}
\usepackage{url}
\usepackage{balance}
\usepackage[hidelinks]{hyperref}

\newcommand{\ci}{\mathrel{\perp\!\!\!\perp}}

\newcommand{\RCCI}{\mathrm{RCCI}}
\newcommand{\ACE}{\mathrm{ACE}}

\begin{document}

\title{Causal Discovery for Resiliency-Prioritized Transmission Expansion Planning
in Power Systems with Grid-Forming Inverter-Based Resources}

%\author{
%\IEEEauthorblockN{Roy McCann}
%\IEEEauthorblockA{Department of Electrical %Engineering & Computer Science, University of Arkansas - Fayetteville\\
%{rmccann@uark.edu}}}

\author{\makebox[\textwidth][c]{%
\begin{tabular}{c}
\IEEEauthorblockN{Roy McCann, Mustafa A. Mhawesh, Sandra Canas Rivera} 
\IEEEauthorblockA{\textit{Department of Electrical Engineering and Computer Science, University of Arkansas} \\
Fayetteville, USA \\
Corresponding author contact: rmccann@uark.edu}
\end{tabular}%
}}

\maketitle

% ---------------------------------------------------------------------------
\begin{abstract}
Transmission planners at regional transmission organizations (RTOs), independent
system operators (ISOs), and NERC-registered entities have limited analytical resources for exploring expansion options that improve resiliency for withstanding and recovering from fault conditions  in electric power systems that are dominated by grid-forming
(GFM) and grid-following (GFL) inverter-based resources (IBRs). Conventional
Existing TPL-001 contingency analysis will assess if a performance benchmark is violated. However, it will not identify which dynamic states and inputs caused the violation, nor how much a protective action will avoid propagation into a cascading outage. This paper proposes an analytical
method that combines time-series causal discovery using the Peter and Clark
Momentary Conditional Independence (PCMCI) algorithm with right/left eigenvector
participation-factor analysis of validated small-signal state-space models. Synthetic
data are generated from linearized models derived from substation phasor measurement
unit (PMU) data, node-breaker topologies under alternative transmission switching
configurations, Heffron--Phillips synchronous-machine models with critical-clearing-time
checks, GFM and GFL inverter controls including phase-locked loops (PLLs) and
fault ride-through logic, and fault transients calibrated from digital fault recorder
(DFR), multi-function recorder, and sequence-of-events recorder (SER) archives.
We show that this construction structurally satisfies the five PCMCI assumptions:
causal sufficiency, realistic stochastic forcing, no contemporaneous causation, a
known maximum lag, and a known functional form. We introduce a Resiliency Causal
Contribution Index (RCCI) that weights PCMCI-derived causal effects on a TPL benchmark
quantity by modal participation. The planning procedure iterates TPL-001-5.1 contingency
categories P0--P7 and extreme events, measures margins to RTO/ISO benchmarks
(transient voltage recovery, oscillatory damping $\zeta\geq 3\%$, PRC-024/PRC-029
ride-through, TPL-008-1 extreme temperature), infers automatic protective
actions, rebuilds the post-action model, and repeats until a cascade is ruled
out or confirmed. When a cascade is likely, the RCCI ranks grid-enhancing technologies, FACTS,
new AC circuits, and HVDC terminals by their ability to interrupt the sequence after
the first contingency. A reproducible simplified case study shows the full
loop. A P1 outage drives the critical electromechanical mode below the 3\% damping
criterion and triggers legacy momentary cessation of a GFL plant. The resulting
thermal overload of the remaining circuit would cascade to loss of the load bus. The RCCI attributes
the violation to the synchronous-machine angle and speed and to the GFM frequency loop,
and correctly predicts that damping-plus-voltage-support alternatives interrupt the
cascade while a dynamic line rating alone does not.
\end{abstract}

% ---------------------------------------------------------------------------
\section{Introduction}
\label{sec:intro}

The bulk power system is changing from one governed by the electromechanical
dynamics of synchronous generators (SGs) to one in which converter controls set
much of the post-disturbance behavior. Grid-following (GFL) IBRs synchronize to the
network through phase-locked loops (PLLs) and act as controlled current sources.
Grid-forming (GFM) IBRs regulate an internal voltage phasor and contribute synchronizing
and damping actions that are partly analogous to those of SGs~\cite{Rocabert2012,Lin2020GFM,NERC2023GFM}.
Several North American disturbances have shown that converter protection and control
logic can turn a routine, normally cleared fault into a large loss of generation.
Examples include momentary cessation, PLL loss of synchronism, and plant-level protective
tripping~\cite{NERC2017BlueCut,NERC2021Odessa,NERC2022Odessa}. These events led to
PRC-028-1, PRC-029-1, and PRC-030-1, together with IEEE Std 2800-2022~\cite{NERC_PRC029,NERC_PRC028,IEEE2800}.

Transmission planners evaluate these risks within the NERC TPL-001-5.1 framework.
It requires steady-state and stability assessments for planning events P0--P7 and
extreme events~\cite{NERC_TPL001}. TPL-008-1 now adds extreme-temperature assessments~\cite{NERC_TPL008}.
RTOs and ISOs add numerical criteria such as minimum post-contingency damping ratios,
transient voltage dip and recovery envelopes, and generator ride-through
requirements. The standard workflow is a pass/fail screen: each contingency is simulated
and each benchmark is checked. A violation triggers a corrective action plan, which is
usually chosen by engineering judgment from a menu of familiar remedies. This workflow
has three gaps for high-IBR systems:
\begin{enumerate}
\item It does not identify the \emph{causal sources} of a violation, that is, which
state variables and exogenous inputs drive the benchmark quantity across the limit.
\item It does not \emph{quantify} each source's contribution in a way that maps to a
class of equipment.
\item It treats the post-contingency topology as final. It does not propagate the
automatic protective actions that a marginal benchmark excursion would trigger, such as
breaker trips, generator trips, inverter momentary cessation, under-voltage load shedding
(UVLS), and remedial action schemes (RAS). These actions are the mechanism by which one
contingency becomes a cascading outage~\cite{Dobson2007,Vaiman2012,Blackout2004}.
\end{enumerate}

Eigenvalue participation factors~\cite{PerezArriaga1982,Kundur1994} answer part
of the first question. They show which states take part in a poorly damped mode, but
participation is associative rather than interventional. A state can participate strongly
in a mode without being the variable whose modification most changes the benchmark
outcome, and exogenous drivers such as stochastic load act as \emph{confounders} that
couple many states at once. Time-series causal discovery offers a complementary,
interventional view. PCMCI~\cite{Runge2019SciAdv,Runge2018Chaos} combines a
PC-style condition-selection stage with a momentary conditional independence (MCI) test.
It recovers lagged causal graphs from high-dimensional, autocorrelated time series with
controlled false-positive rates, and it has been applied widely in the Earth sciences~\cite{Runge2019NatComm}.
PCMCI, however, is valid only under strong assumptions: causal sufficiency, no
contemporaneous effects at the sampling interval, a known maximum lag, faithfulness, and
stationarity. Raw field data from a transmission system rarely satisfy these assumptions.

\textbf{Key Concept:} PCMCI is not directly applied to power system operational measurements. Instead, we apply PCMCI to synthetic data generated from \emph{operationally validated} small-signal
state-space models, built from substation PMU data, node-breaker configurations, and
recorder-calibrated fault transients. Because the generating mechanism is a known
zero-order-hold (ZOH) discretized linear system driven by measured load variability,
each PCMCI assumption can be established by construction rather than assumed.
The same state-space model supplies the left and right eigenvectors needed for participation
factors. These factors are used both to screen for confounders and to weight causal effects
into a contribution metric for each benchmark.

\textbf{Contributions.}
\begin{enumerate}
\item A formal mapping, given in Table~\ref{tab:assumptions}, from the five PCMCI
assumptions to properties of PMU-validated, ZOH-discretized small-signal models.
The mapping makes PCMCI applicable to planning studies.
\item A small-signal model formulation for hybrid SG/GFM/GFL systems. It includes
Heffron--Phillips SGs with critical-clearing-time checks, PLL dynamics with a weak-grid
algebraic loop, and fault-ride-through and momentary-cessation logic triggered by
recorder-calibrated fault-equivalent inputs.
\item The \emph{Resiliency Causal Contribution Index} (RCCI). For each TPL benchmark, it
combines PCMCI-restricted average causal effects with modal participation factors, and it
uses input participation to flag confounders.
\item An iterative, resiliency-prioritized transmission expansion planning (TEP)
procedure (Fig.~\ref{fig:flow}, Algorithm~\ref{alg:rptep}). It propagates protective
actions through rebuilt post-action models to detect probable cascades, and it uses the
RCCI to rank grid-enhancing technologies (GETs), FACTS, new AC circuits, and HVDC terminals
by their ability to interrupt the cascade after the first contingency.
\item A reproducible simplified case study that exercises every step of the procedure.
\end{enumerate}

Prior work in causal discovery for electric power systems has approached problems with cascading failures and forecasting electricity demand. PCMCI was used in~\cite{Mhawesh2026} in the context of causal inference to determine the contributing factors to fault propagation in electric power systems. Causal detection models are developed in ~\cite{Maisonnave2024} for predicting future energy demand. A causal framework is developed in ~\cite{Caetano2026} to analyze cascading failures through a data-derived spatiotemporal Structural Causal Model. Online transient stability methods based  on causal inference are developed ~\cite{Shahbaz2026}. These previous research efforts have not accounted for transmission system design and the distinctive behavior of inverter-based generation. In this paper, we provide a new method for engineering personnel to assess options for increasing power system resiliency when evaluating design and equipment options throughout the transmission expansion planning process.

% ---------------------------------------------------------------------------
\section{Background and Related Work}
\label{sec:background}

\subsection{Cascading outages, resilience, and planning}
Cascading failure research spans topological, quasi-steady-state (e.g., OPA), and
dynamic models~\cite{Dobson2007,Vaiman2012}. Post-event analyses of the 2003 Northeast
blackout~\cite{Blackout2004} and of IBR-related disturbances~\cite{NERC2017BlueCut,NERC2021Odessa,NERC2022Odessa}
show that the decisive step is often an automatic protective action whose trigger
lies close to its setting. Resilience frameworks highlight the ability to withstand,
absorb, and recover~\cite{Panteli2015}. TEP formulations have historically optimized cost
under adequacy and security constraints~\cite{Hemmati2013}. Recent FERC actions, Orders
No.~881 and No.~1920, require ambient-adjusted line ratings and long-term regional planning
that considers GETs such as dynamic line ratings, advanced power-flow control, and
transmission switching~\cite{FERC881,FERC1920}. Optimal transmission switching~\cite{Fisher2008}
shows that topology is itself a decision variable, which motivates our use of node-breaker
models.

\subsection{Small-signal stability with IBRs}
Small-signal modeling of SGs with exciters dates to Heffron and Phillips~\cite{Heffron1952}
and deMello and Concordia~\cite{deMello1969}. Modal analysis with participation factors
is standard practice~\cite{PerezArriaga1982,Kundur1994}, with extensions that separate
state and input participation~\cite{Hashlamoun2009}. Inverter small-signal models include
droop-controlled GFM converters~\cite{Pogaku2007} and GFL converters, whose PLL-based
synchronization becomes fragile in weak grids~\cite{Wang2020Sync}. The extended stability
classification~\cite{Hatziargyriou2021} adds converter-driven and resonance categories.
Ambient PMU data support modal estimation for model validation~\cite{Hauer1990,C37118}.

\subsection{Causal discovery for time series}
Granger causality~\cite{Granger1969} tests whether the past of one variable improves
prediction of another. It is biased by autocorrelation and by unobserved common drivers.
Constraint-based methods built on conditional independence~\cite{Spirtes2000,Pearl2009}
underlie PCMCI~\cite{Runge2019SciAdv}. Linear mediation analysis on the discovered graph
gives path-specific and total causal effects~\cite{Runge2015NatComm}. We are not aware of prior work that
links PCMCI to TPL planning benchmarks through participation factors.

% ---------------------------------------------------------------------------
\section{Small-Signal Models and Synthetic Data}
\label{sec:models}

\subsection{Node-breaker topology and switching configurations}
Each planning case starts from a node-breaker model of the RTO footprint. A topology
processor maps the breaker and switch status vector $\mathbf{s}\in\{0,1\}^{n_b}$ to a
bus-branch network with admittance matrix $\mathbf{Y}(\mathbf{s})$. Alternative transmission
switching configurations, such as split buses, bus-tie open or closed, and breaker-and-a-half
paths, are enumerated as a set $\mathcal{S}$. This matters because the same P2
(bus-section) or P4/P5 (stuck breaker, relay failure) event isolates different elements
under different $\mathbf{s}$. Every contingency and protective action in
Section~\ref{sec:tep} is applied as a change to $\mathbf{s}$, never directly to the
bus-branch model.

\subsection{Linearized hybrid SG/GFM/GFL model}
For a given $\mathbf{s}$ and operating point, the differential-algebraic model is
linearized and the algebraic network variables are eliminated (Kron reduction):
\begin{align}
\dot{\mathbf{x}} &= \mathbf{A}(\mathbf{s})\mathbf{x} + \mathbf{B}(\mathbf{s})\mathbf{u}, &
\mathbf{y} &= \mathbf{C}(\mathbf{s})\mathbf{x} + \mathbf{D}(\mathbf{s})\mathbf{u},
\label{eq:ss}
\end{align}
where $\mathbf{x}$ collects device states, $\mathbf{u}$ collects exogenous inputs
(load active and reactive power, and fault-equivalent inputs), and $\mathbf{y}$ collects
the algebraic benchmark quantities: bus voltages, branch flows, and PLL angle errors.

\textit{Synchronous generators.} Each SG is represented by the Heffron--Phillips model~\cite{Heffron1952,deMello1969}
with a static exciter and a first-order governor/turbine:
\begin{align}
\Delta\dot\delta &= \omega_0\Delta\omega, \quad
2H\Delta\dot\omega = \Delta P_m - \Delta P_e - D\Delta\omega, \nonumber\\
T'_{d0}\Delta\dot E'_q &= -\Delta E'_q/K_3 - K_4\Delta\delta_{L} + \Delta E_{fd},
\label{eq:hp}\\
T_A\Delta\dot E_{fd} &= -\Delta E_{fd} - K_A\Delta V_t, \quad
\Delta V_t = K_5\Delta\delta_{L} + K_6\Delta E'_q, \nonumber
\end{align}
where $\Delta\delta_L$ is the rotor angle relative to the connecting bus. The electrical
power $\Delta P_e$ comes from the reduced network (it plays the role of $K_1\Delta\delta+K_2\Delta E'_q$).
The constant $K_5$ becomes more negative as the external system weakens, which is the
classical mechanism by which high-gain excitation erodes electromechanical damping~\cite{deMello1969}.
Transient-stability adequacy is checked using the critical clearing time. For a bolted
fault with $P_e=0$ during the fault and post-fault $P_{\max}$ equal to its pre-fault value,
\begin{equation}
\delta_{cr}=\cos^{-1}\!\Big[\tfrac{P_m}{P_{\max}}(\pi-2\delta_0)-\cos\delta_0\Big],\quad
t_{cr}=\sqrt{\tfrac{4H(\delta_{cr}-\delta_0)}{\omega_0P_m}},
\label{eq:cct}
\end{equation}
and is compared against the clearing times measured by DFRs for the protection scheme
involved (normal clearing for P1--P3, and backup or breaker-failure clearing for P4--P5)~\cite{Kundur1994}.
In RTO-scale studies, \eqref{eq:cct} serves as a screen, and nonlinear time-domain
simulation gives the definitive result.

\textit{Grid-forming IBRs.} Each GFM is represented as a droop-controlled voltage source
with low-pass power measurement~\cite{Pogaku2007}:
\begin{equation}
\begin{aligned}
&\dot\theta_g=\omega_0\Delta\omega_g,\qquad T_f\Delta\dot\omega_g=-\Delta\omega_g-m_p\Delta P_g,\\
&T_v\Delta\dot V_g=-\Delta V_g-n_q\Delta Q_g .
\end{aligned}
\label{eq:gfm}
\end{equation}
Virtual-synchronous-machine variants replace \eqref{eq:gfm} with a swing equation that has
programmable inertia and damping. Current limiting switches the GFM to a current-source
representation and is treated as a protective action (Section~\ref{sec:tep}).

\textit{Grid-following IBRs with PLL and ride-through logic.} The GFL plant includes a
synchronous-reference-frame PLL with a phase-detector filter, a PI loop filter, and a
first-order inner current loop with fast frequency response:
\begin{align}
T_m\dot e_f &= e-e_f,\quad \dot\theta_{pll}=K_pe_f+x_{pll},\quad \dot x_{pll}=K_ie_f, \nonumber\\
T_i\Delta\dot i_d &= -\Delta i_d - K_f\,\omega_{pll}/\omega_0,
\label{eq:gfl}
\end{align}
where $e=\theta_L-\theta_{pll}$ is the PLL angle error. Because the current is injected at
the PLL angle, the active power is $\Delta P_{gfl}\approx\Delta i_d+\gamma e$. With Thevenin
synchronizing strength $s_{th}$ seen by the plant, the load-bus angle then satisfies
\begin{equation}
\theta_L=\frac{b_1\delta_r-\Delta P_L+\Delta i_d-\gamma\theta_{pll}}{s_{th}-\gamma},
\label{eq:loop}
\end{equation}
which shows the familiar weak-grid limit $s_{th}\to\gamma$, where the PLL loop gain grows
without bound~\cite{Wang2020Sync}. Ride-through behavior follows the plant's as-built
settings and is checked against PRC-029-1 and IEEE 2800~\cite{NERC_PRC029,IEEE2800}. For
legacy units, momentary cessation is entered when the terminal voltage drops below
$V_{MC}$ or the PLL angle jumps by more than a threshold. PRC-029-1 references 25 electrical
degrees for non-fault phase jumps. Output is then restored after a recovery delay or not
at all. These discrete actions are \emph{not} part of \eqref{eq:ss}. They are inferred from
benchmark trajectories and applied as topology or parameter changes in the next iteration.

\subsection{Validation against PMU data}
The model in \eqref{eq:ss} is accepted for a topology $\mathbf{s}$ only after it is checked
against ambient and ringdown PMU records (IEEE C37.118~\cite{C37118}) collected at the
transmission substations, for operating periods with the same breaker status. The checks
are as follows:
\begin{enumerate}
\item The frequency and damping of electromechanical modes estimated from ambient data by
Prony or subspace methods~\cite{Hauer1990} must match $\mathrm{eig}(\mathbf{A})$ within
tolerance.
\item Mode shapes from cross-spectra must match the right eigenvectors.
\item Simulated voltage and frequency responses to recorded events must match the PMU
trajectories.
\end{enumerate}
Mismatches are corrected by adjusting uncertain control parameters, such as PLL gains,
droop constants, and exciter gains, within manufacturer ranges.

\subsection{Fault-equivalent inputs calibrated from recorders}
A small-signal model cannot represent a bolted fault exactly. We therefore represent each
fault class by a \emph{fault-equivalent input} $u_f(t)$ applied through calibrated
input vectors. These vectors produce a reactive drain at the faulted bus and electrical
power deficits at nearby machines. The magnitudes and the clearing duration of the pulse
are fitted to a repository of historical DFR, multi-function recorder, and SER records
for the same voltage class and protection scheme. PRC-028-1 now requires comparable records
from IBR plants~\cite{NERC_PRC028}. Because the voltage dip at the faulted bus then depends
on local voltage stiffness, as it does in the field, GETs and FACTS devices change both the
depth of the dip and the resulting protective response.

\subsection{Synthetic data generation}
For PCMCI, \eqref{eq:ss} is discretized with a ZOH at the PMU reporting interval
$\Delta t$ (e.g., 30 or 60 frames/s):
\begin{equation}
\mathbf{x}_{k+1}=\mathbf{A}_d\mathbf{x}_k+\mathbf{B}_d\mathbf{u}_k,\quad
\mathbf{A}_d=e^{\mathbf{A}\Delta t},\ \mathbf{B}_d=\!\int_0^{\Delta t}\!e^{\mathbf{A}\tau}d\tau\,\mathbf{B}.
\label{eq:zoh}
\end{equation}
The load inputs $\mathbf{u}_k$ are taken from PMU records of feeder and transformer loading
under nominal stochastic variability (no events), detrended and scaled to the planning case.
Measurement noise consistent with C37.118 total-vector-error limits is added to the outputs.
The data set $\mathbf{X}=[\mathbf{x}_k^\top\ \mathbf{u}_k^\top]_{k=1}^T$ contains every
dynamic state and every exogenous input. Algebraic quantities, which are contemporaneous
functions of $\mathbf{x}_k$ and $\mathbf{u}_k$, are deliberately \emph{excluded} from the
causal-discovery set and are recovered through $\mathbf{C}$ and $\mathbf{D}$.

% ---------------------------------------------------------------------------
\section{Applicability of PCMCI}
\label{sec:pcmci}

\subsection{The PCMCI algorithm}
Let $X^j_t$ denote variable $j$ at time $t$ and let $\tau_{\max}$ be the maximum lag. PCMCI~\cite{Runge2019SciAdv}
proceeds in two stages.

\textit{PC$_1$ condition selection.} For each target $X^j_t$, start from all lagged
candidates $\{X^i_{t-\tau}\}_{i,\,1\le\tau\le\tau_{\max}}$ and iteratively remove each
candidate $X^i_{t-\tau}$ for which
$X^i_{t-\tau}\ci X^j_t\mid\mathcal{S}$ at level $\alpha_{PC}$. Here $\mathcal{S}$ contains the
$p$ strongest remaining candidates, and $p$ is increased until no further removals occur.
This yields a superset of parents $\widehat{\mathcal{P}}(X^j_t)$.

\textit{MCI test.} For every pair $(X^i_{t-\tau},X^j_t)$, test
\begin{equation}
X^i_{t-\tau}\ci X^j_t\ \Big|\ \widehat{\mathcal{P}}(X^j_t)\setminus\{X^i_{t-\tau}\},\
\widehat{\mathcal{P}}(X^i_{t-\tau}),
\label{eq:mci}
\end{equation}
with the partial-correlation statistic
$I^{MCI}_{i\to j}(\tau)=\rho(X^i_{t-\tau},X^j_t\mid\mathcal{Z})$. Conditioning on the
parents of the \emph{cause} removes autocorrelation effects that inflate false positives in
Granger-type tests. The $p$-values are adjusted by the Benjamini--Hochberg false discovery
rate (FDR) procedure~\cite{Benjamini1995}, and links with $q\le\alpha$ form the estimated
causal graph $\widehat{\mathcal{G}}$.

\subsection{Assumptions satisfied by construction}
Table~\ref{tab:assumptions} lists the five conditions identified in the problem statement,
together with faithfulness and stationarity, and shows how each one follows from the data
construction of Section~\ref{sec:models}.

\begin{table}[t]
\centering
\caption{PCMCI assumptions and how validated small-signal synthetic data satisfy them.}
\label{tab:assumptions}
\footnotesize
\begin{tabular}{@{}p{0.27\columnwidth}p{0.66\columnwidth}@{}}
\toprule
Assumption & Satisfied by construction because\ldots\\
\midrule
(1) Causal sufficiency (all confounders observed) &
$\mathbf{X}$ contains every state of \eqref{eq:ss} and every exogenous input $\mathbf{u}$.
Unmodeled common drivers are screened by input participation factors (Sec.~\ref{sec:rcci}),
which flag any input that excites several states of a critical mode.\\[2pt]
(2) Realistic stochastic forcing &
$\mathbf{u}_k$ is measured PMU load variability under nominal conditions, not white noise
of arbitrary color. The excitation therefore matches the spectrum the real system sees.\\[2pt]
(3) No contemporaneous causation &
Under ZOH \eqref{eq:zoh}, $\mathbf{x}_{k+1}$ depends only on $(\mathbf{x}_k,\mathbf{u}_k)$, so
every cause strictly precedes its effect by $\ge 1$ sample. Algebraic (instantaneous)
variables are excluded from $\mathbf{X}$.\\[2pt]
(4) Known maximum lag &
A fully observed ZOH system is first-order Markov, so the true lag is $\tau=1$.
$\tau_{\max}$ is set to a small value ($2$--$4$) only to absorb measurement noise and the
color of $\mathbf{u}$. It is not a guess about unknown physics.\\[2pt]
(5) Known functional form &
The state-space model is linear in deviations, so linear partial correlation (ParCorr) is
the correct conditional-independence test, and the fitted parent model is a VAR
restricted to $\widehat{\mathcal{G}}$.\\[2pt]
Faithfulness, stationarity &
Stationarity holds for each fixed $\mathbf{s}$ when $\mathbf{A}$ is Hurwitz. Faithfulness
can fail only for parameter-specific cancellations, which are checked by comparing
$\widehat{\mathcal{G}}$ with the sparsity of $[\mathbf{A}_d\,\mathbf{B}_d]$.\\
\bottomrule
\end{tabular}
\end{table}

An important corollary is that the causal claims made here concern the \emph{validated
model} of the system. They inherit the model's fidelity, which is why PMU validation is a
prerequisite step and not an optional one. The value of running PCMCI on data generated
by a known model, rather than reading $\mathbf{A}_d$ directly, is threefold:
\begin{enumerate}
\item It gives a sparse causal graph at the PMU timescale, with statistical significance,
that is robust to measurement noise. The matrix $\mathbf{A}_d$ is dense because of
multi-path fill-in.
\item It yields effect sizes standardized by the variability that loads actually induce.
\item It lets planners use the same pipeline on field PMU data where model coverage is
incomplete, with the synthetic results serving as a reference graph.
\end{enumerate}

\subsection{Average causal effect on a benchmark quantity}
Let $\widehat{\boldsymbol{\Phi}}_\tau$ denote the standardized lag-$\tau$ coefficients of
the linear model restricted to the parents in $\widehat{\mathcal{G}}$, and let
$\boldsymbol{\Psi}_h=\sum_{\tau=1}^{\min(h,\tau_{\max})}\widehat{\boldsymbol{\Phi}}_\tau\boldsymbol{\Psi}_{h-\tau}$, with $\boldsymbol{\Psi}_0=\mathbf{I}$,
be its impulse responses~\cite{Runge2015NatComm}. For a benchmark output
$y_b=\mathbf{c}_b^\top\mathbf{x}+\mathbf{d}_b^\top\mathbf{u}$, define the average causal
effect of a one-standard-deviation intervention on variable $i$ as
\begin{equation}
\ACE_{i\to b}=\Delta t\sum_{h=1}^{H}\big|\,\mathbf{w}_b^\top\boldsymbol{\Psi}_h\mathbf{e}_i\big|,\qquad
\mathbf{w}_b=\boldsymbol{\Sigma}[\mathbf{c}_b;\mathbf{d}_b],
\label{eq:ace}
\end{equation}
where $\boldsymbol{\Sigma}$ is the diagonal matrix of standard deviations and $H\Delta t$
spans several periods of the slowest mode of interest.

% ---------------------------------------------------------------------------
\section{Participation Factors, Confounder Screening, and the RCCI}
\label{sec:rcci}

Let $\mathbf{A}\mathbf{v}_k=\lambda_k\mathbf{v}_k$ and $\mathbf{w}_k^\top\mathbf{A}=\lambda_k\mathbf{w}_k^\top$,
with $\mathbf{w}_k^\top\mathbf{v}_k=1$. The normalized participation of state $i$ in mode $k$ is~\cite{PerezArriaga1982}
\begin{equation}
p_{ik}=\frac{|v_{ik}w_{ki}|}{\sum_{\ell}|v_{\ell k}w_{k\ell}|},\qquad
\zeta_k=\frac{-\Re\lambda_k}{|\lambda_k|}.
\label{eq:pf}
\end{equation}

\textit{Input participation and confounder screening.} Following the input--state
distinction of~\cite{Hashlamoun2009}, the modal excitation share of input $j$ in mode $k$ is
\begin{equation}
\eta_{jk}=\frac{|\mathbf{w}_k^\top\mathbf{b}_j|\,\|\mathbf{v}_k\|}{\sum_{\ell}|\mathbf{w}_k^\top\mathbf{b}_\ell|\,\|\mathbf{v}_k\|}.
\label{eq:eta}
\end{equation}
An input $u_j$ is flagged as a \emph{confounder} of a critical mode $k^\star$ if two
conditions hold: (i) $\eta_{jk^\star}$ is non-negligible, and (ii) PCMCI finds lagged links
from $u_j$ into two or more of the states with the largest $p_{ik^\star}$. Omitting a flagged
input from $\mathbf{X}$ violates causal sufficiency and produces spurious state-to-state
links. Section~\ref{sec:case} demonstrates this. In field-data applications, a confounder
flag signals that the corresponding measurement must be added before causal conclusions
are drawn.

\textit{Resiliency Causal Contribution Index.} For benchmark $b$ violated, or nearly
violated, through critical mode $k^\star$, we define
\begin{equation}
\RCCI_{i}^{(b)}=\frac{p_{ik^\star}\,\ACE_{i\to b}}{\sum_{\ell\in\mathcal{X}}p_{\ell k^\star}\,\ACE_{\ell\to b}},\qquad i\in\mathcal{X},
\label{eq:rcci}
\end{equation}
where $\mathcal{X}$ is the index set of state variables. The participation factor localizes
the contribution to the mode that carries the violation. The ACE adds the data-driven,
interventional strength with which each state drives the benchmark quantity in the
post-contingency causal graph. Exogenous inputs are reported separately through
$\eta_{jk^\star}$ because their scale is not commensurate with state participation. Since
$\sum_i\RCCI_i^{(b)}=1$, the index gives the planner a budget-like allocation of
\emph{causal responsibility}, which maps onto equipment classes as shown in
Table~\ref{tab:map}.

\begin{table}[t]
\centering
\caption{Mapping dominant RCCI state classes to candidate expansion equipment.}
\label{tab:map}
\footnotesize
\begin{tabular}{@{}p{0.36\columnwidth}p{0.57\columnwidth}@{}}
\toprule
Dominant RCCI class & Candidate remedies (primary mechanism)\\
\midrule
SG rotor angle $\Delta\delta$; corridor angle & New AC circuit, series compensation, HVDC
(synchronizing strength, transfer path)\\
SG speed $\Delta\omega$, GFM frequency $\Delta\omega_g$ & POD on STATCOM/SVC or HVDC, GFM
damping retuning, PSS (damping torque)\\
Excitation $\Delta E'_q$, $\Delta E_{fd}$ & PSS retuning, exciter gain revision\\
PLL states $\Delta\theta_{pll},\Delta x_{pll},\Delta e_f$ & GFM conversion, synchronous
condenser, STATCOM (grid strength, voltage stiffness)\\
GFM/GFL voltage states & STATCOM, SVC, synchronous condenser\\
Post-action thermal flow (from $\mathbf{y}$) & Short-duration DLR, power-flow control, RAS\\
\bottomrule
\end{tabular}
\end{table}

% ---------------------------------------------------------------------------
\section{Resiliency-Prioritized TEP Procedure}
\label{sec:tep}

\begin{figure}[t]
\centering
\resizebox{\columnwidth}{!}{%
\begin{tikzpicture}[
  node distance=3.2mm and 4mm,
  box/.style={draw, rounded corners=1.5pt, fill=blue!4, align=center, font=\scriptsize,
              text width=40mm, minimum height=6.5mm, inner sep=2pt},
  dec/.style={draw, diamond, aspect=2.6, fill=orange!8, align=center, font=\scriptsize,
              inner sep=0.6pt, text width=19mm},
  side/.style={draw, rounded corners=1.5pt, fill=green!5, align=center, font=\scriptsize,
              text width=27mm, minimum height=6.5mm, inner sep=2pt},
  arr/.style={-{Stealth[length=1.6mm]}, thick}]
\node[box] (n1) {1. Node-breaker case $\mathbf{s}\in\mathcal{S}$, SG + GFM + GFL dispatch};
\node[box, below=of n1] (n2) {2. Linearize \& validate vs.\ PMU modes/events $\Rightarrow (\mathbf{A},\mathbf{B},\mathbf{C},\mathbf{D})$};
\node[box, below=of n2] (n3) {3. Synthetic data (ZOH, PMU load variability); PCMCI $\Rightarrow\widehat{\mathcal{G}}$; participation $p_{ik},\eta_{jk}$};
\node[box, below=of n3] (n4) {4. TPL-001-5.1 P0--P7 + extreme events (DFR-calibrated $u_f$)};
\node[box, below=of n4] (n5) {5. Benchmark margins: $V$ dip/recovery, $\zeta\!\ge\!3\%$, PRC-024/029 ride-through, TPL-008-1};
\node[box, below=of n5] (n6) {6. Causal sources: $\RCCI_i^{(b)}$ and confounder flags};
\node[box, below=of n6] (n7) {7. Infer protective actions (breaker/generator trip, momentary cessation, UVLS, RAS, GFM current limit)};
\node[dec, below=4mm of n7] (d1) {Further actions?};
\node[side, right=5mm of n4] (r1) {8. Update $\mathbf{s}$ and parameters; rebuild post-action model};
\node[dec, below=4mm of d1] (d2) {Cascade probable?};
\node[side, right=5mm of d2, yshift=0mm] (r2) {10. Screen GETs, FACTS, AC, HVDC ranked by RCCI (Table~\ref{tab:map}); re-run from step 2};
\node[box, below=4mm of d2, fill=gray!8] (n9) {9. Report margins, causal sources, accepted plan};
\foreach \a/\b in {n1/n2,n2/n3,n3/n4,n4/n5,n5/n6,n6/n7,n7/d1} \draw[arr] (\a)--(\b);
\draw[arr] (d1.east) -| node[pos=0.25, above, font=\scriptsize]{yes} (r1.south);
\draw[arr] (r1.north) |- (n2.east);
\draw[arr] (d1) -- node[right, font=\scriptsize]{no} (d2);
\draw[arr] (d2) -- node[above, font=\scriptsize]{yes} (r2);
\draw[arr] (d2) -- node[right, font=\scriptsize]{no} (n9);
\draw[arr] (r2.north) -- ++(0,6mm) -| ($(r1.east)+(3mm,0)$) |- ($(n2.east)+(0,1.2mm)$);
\end{tikzpicture}}
\caption{Resiliency-prioritized transmission expansion planning procedure. Steps 2--8
iterate over post-action topologies until no further automatic action is triggered. Step 10
evaluates expansion alternatives that interrupt a probable cascade after the first contingency.}
\label{fig:flow}
\end{figure}
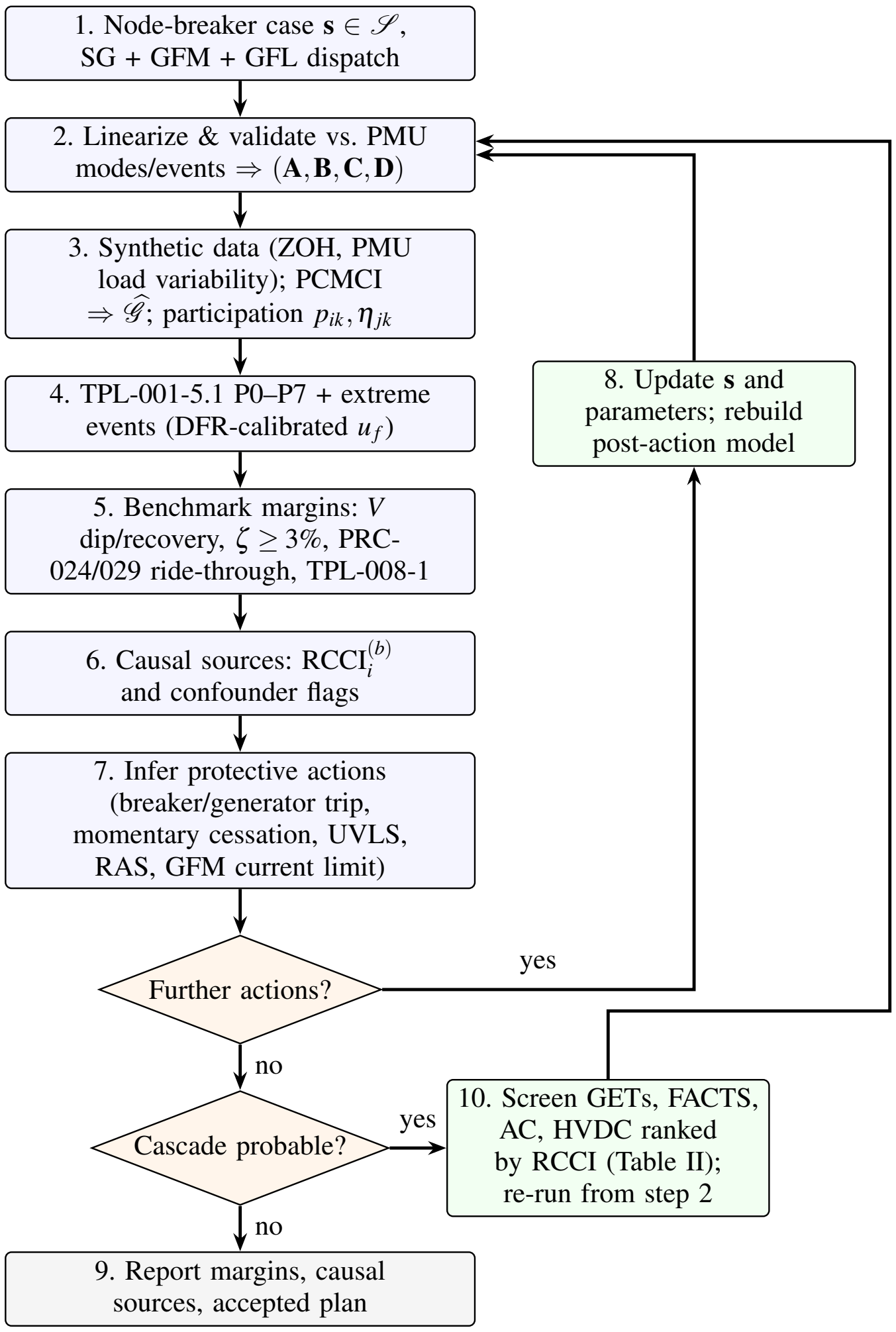

\subsection{Benchmarks and margins}
For each contingency $c$ in TPL-001-5.1 P0--P7 and the extreme events~\cite{NERC_TPL001},
the procedure computes signed margins. A positive margin means the criterion is met, and
a margin near zero means the case is close to the limit. The thresholds below are typical
RTO/ISO values. Each entity substitutes its own published criteria.

\textit{(1) Transient voltage dip and recovery.} For load-serving BES buses,
$m_{V,\mathrm{rec}}=V(t_c+20\,\mathrm{s})/V_{\mathrm{pre}}-0.80$, which guards against
induction-motor stalling and fault-induced delayed voltage recovery. A second margin,
$m_{V,\mathrm{dip}}=T_{\mathrm{allow}}-t\{V<V_{\mathrm{dip}}\}$, applies a dip threshold
$V_{\mathrm{dip}}\in[0.70,0.80]$~p.u. sustained for no more than $T_{\mathrm{allow}}$ of a few seconds.

\textit{(2) Oscillatory damping.} $m_\zeta=(\min_k\zeta_k-0.03)/0.03$ over post-contingency
oscillatory modes, evaluated both from $\mathbf{A}(\mathbf{s}_c)$ and, as a cross-check,
from ringdown fits of simulated trajectories.

\textit{(3) Ride-through.} For SGs, the distance of the terminal voltage and frequency
trajectories to the PRC-024 relay-setting envelope. For IBRs, the distance to the PRC-029-1
``must ride-through'' zones and the PLL phase-jump excursion relative to the as-built
trip or momentary-cessation threshold~\cite{NERC_PRC029,IEEE2800}. The CCT margin
$t_{cr}-t_{\mathrm{clear}}$ from \eqref{eq:cct} is also reported.

\textit{(4) Extreme temperature.} The TPL-008-1 benchmark cases~\cite{NERC_TPL008} are
re-run with temperature-dependent line ratings, load levels, and resource availability.
The same margins are then evaluated.

\subsection{Protective-action inference and topology update}
For each margin that is negative or below a planning tolerance, the trajectories of
$\mathbf{y}$ are compared against the protection settings in the RTO's model data:
distance and overcurrent elements and thermal overload timers on lines, generator
out-of-step, under/over-frequency and voltage relays, IBR momentary-cessation and trip
logic, GFM current limits, UVLS/UFLS stages, and RAS arming conditions. Each triggered
action is mapped to a change in $\mathbf{s}$ (breaker opening) or in device parameters
(e.g., setting a GFL plant's online flag to zero, or a GFM switching to current-limited
mode). A new linearized model is then built around the post-action equilibrium. PCMCI and
the participation analysis are repeated on that model (Fig.~\ref{fig:flow}, step 8), so the
causal graph tracks how the system is reconfigured during the cascade.

\subsection{Cascade criterion and expansion screening}
A contingency is labeled a \emph{probable cascade} if the iteration produces two or more
consecutive automatic actions beyond the initiating event, or reaches non-consequential
load loss, islanding, or uncontrolled separation, all of which TPL-001-5.1 prohibits for the
event category concerned. Candidate expansions are then generated from Table~\ref{tab:map}
in RCCI order and evaluated by re-running the loop from step 2. The preferred alternatives
are those that (a) restore all margins after the \emph{first} contingency, so that no
protective action is triggered, and (b) do so at least cost. Algorithm~\ref{alg:rptep}
summarizes the procedure.

\begin{algorithm}[t]
\caption{Resiliency-prioritized TEP with PCMCI and RCCI}
\label{alg:rptep}
\footnotesize
\begin{algorithmic}[1]
\For{each base topology $\mathbf{s}_0\in\mathcal{S}$ and dispatch}
  \State Build and PMU-validate $(\mathbf{A},\mathbf{B},\mathbf{C},\mathbf{D})(\mathbf{s}_0)$
  \For{each contingency $c\in\{\mathrm{P0},\dots,\mathrm{P7},\mathrm{EE}\}$}
    \State $\mathbf{s}\gets\mathbf{s}_0\oplus c$;\ $\mathcal{A}\gets\emptyset$ \Comment{actions so far}
    \Repeat
      \State Generate $\mathbf{X}$ via \eqref{eq:zoh}; run PCMCI $\Rightarrow\widehat{\mathcal{G}}(\mathbf{s})$
      \State Compute $\lambda_k,p_{ik},\eta_{jk}$; benchmark margins $m_b$
      \State $\RCCI^{(b)}\gets$ \eqref{eq:rcci} for each $b$ with $m_b<\epsilon_b$
      \State $\mathcal{A}_{\mathrm{new}}\gets$ protective actions implied by $\mathbf{y}(t)$
      \State $\mathbf{s}\gets\mathbf{s}\oplus\mathcal{A}_{\mathrm{new}}$;\ $\mathcal{A}\gets\mathcal{A}\cup\mathcal{A}_{\mathrm{new}}$
    \Until{$\mathcal{A}_{\mathrm{new}}=\emptyset$ \textbf{or} load loss/islanding}
    \If{$|\mathcal{A}|\ge 2$ \textbf{or} load loss/islanding} \Comment{probable cascade}
      \State Rank equipment classes by $\RCCI$ (Table~\ref{tab:map}); test each
      \State Keep alternatives with $m_b\ge\epsilon_b\ \forall b$ and $\mathcal{A}=\emptyset$ after $c$
    \EndIf
  \EndFor
\EndFor
\end{algorithmic}
\end{algorithm}

% ---------------------------------------------------------------------------
\section{Example Case Study}
\label{sec:case}

\subsection{Test system and setup}
To show every step reproducibly, we use a deliberately simplified system rather than
an complete RTO-scale model. An SG (0.45~p.u. dispatch, 100~MVA base) feeds a load bus over line L1. A GFM
battery plant (0.25~p.u.) feeds the same bus over a two-circuit corridor L2a/L2b. A GFL
solar plant (0.30~p.u.) is connected at the load bus, which serves 1.0~p.u. of load. The
resulting instantaneous IBR share is 55\%. The model has 11 states:
$\Delta\delta,\Delta\omega,\Delta E'_q,\Delta E_{fd},\Delta P_m$ (SG),
$\Delta\omega_g,\Delta V_g$ (GFM), and $\Delta\theta_{pll},\Delta x_{pll},\Delta i_d,\Delta e_f$ (GFL).
It has two load inputs, $\Delta P_L$ and $\Delta Q_L$, and one fault-equivalent input $u_f$.
Static and short-duration DLR ratings are 0.75/0.90~p.u. for L1 and 0.35/0.45~p.u. per L2
circuit.

Parameters were chosen to be physically plausible and to exhibit the mechanisms of interest.
They are \emph{not} fitted to a specific utility system. In particular, in place of real PMU
load records and a DFR archive, we use (i) a two-time-constant Ornstein--Uhlenbeck surrogate
for load variability (time constants of 8~s and 0.5~s) and (ii) a fault-equivalent input with
a 5-cycle normal-clearing duration and 15-cycle backup clearing. The legacy GFL plant
enters momentary cessation without recovery when its terminal voltage falls below $V_{MC}=0.50$~p.u.,
or when the PLL angle error exceeds $25^\circ$. Synthetic PMU data consist of
$T=18{,}000$ samples (10~min at 30~frames/s) with 1\% measurement noise. PCMCI uses ParCorr,
$\tau_{\max}=3$, $\alpha_{PC}=0.05$, and FDR-adjusted $\alpha=0.01$. We implemented PCMCI
directly from~\cite{Runge2019SciAdv}. Each run takes about 30~s in single-threaded Python.

\subsection{Iteration 1: P1 outage of L2a}
\begin{figure}[t]
\centering
\includegraphics[width=\columnwidth]{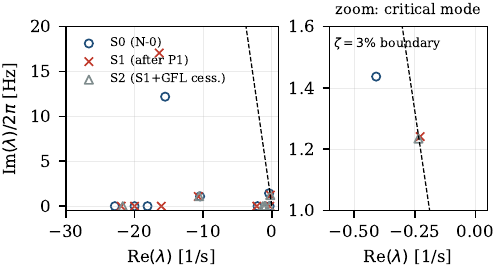}
\caption{Eigenvalues for S0, S1, and S2 (left) and zoom on the critical electromechanical
mode with the $\zeta=3\%$ boundary (right).}
\label{fig:eig}
\end{figure}

Table~\ref{tab:modes} and Fig.~\ref{fig:eig} compare the oscillatory modes. In the N-0 case
S0, the least-damped mode is a 1.44~Hz electromechanical mode dominated by $\Delta\delta$,
$\Delta\omega$, and $\Delta\omega_g$, with $\zeta=4.53\%$ ($m_\zeta=+51\%$). A
12.2~Hz PLL mode is well damped. A three-phase fault on L2a cleared in 5 cycles (P1) leaves
the single circuit L2b (case S1). The weaker corridor reduces the synchronizing coefficient,
makes $K_5$ more negative, and lowers the critical mode to $\zeta=2.92\%$. This is a
\emph{violation} of the 3\% criterion ($m_\zeta=-2.7\%$). It is a marginal one that a pass/fail
screen would flag without explaining.

\begin{table}[t]
\centering
\caption{Oscillatory modes (dominant states by participation).}
\label{tab:modes}
\footnotesize
\setlength{\tabcolsep}{3pt}
\begin{tabular}{@{}lccccl@{}}
\toprule
Case & $\Re\lambda$ [1/s] & $f$ [Hz] & $\zeta$ [\%] & $m_\zeta$ & Dominant states\\
\midrule
S0 (N-0) & $-0.409$ & 1.437 & 4.53 & $+0.51$ & $\Delta\delta,\Delta\omega,\Delta\omega_g$\\
 & $-15.55$ & 12.19 & 19.9 & --- & $\Delta e_f,\Delta\theta_{pll},\Delta x_{pll}$\\
S1 (P1: L2a out) & $-0.228$ & 1.242 & \textbf{2.92} & $\mathbf{-0.03}$ & $\Delta\delta,\Delta\omega,\Delta\omega_g$\\
 & $-16.41$ & 17.06 & 15.1 & --- & $\Delta e_f,\Delta\theta_{pll},\Delta x_{pll}$\\
S2 (S1 + GFL cess.) & $-0.235$ & 1.235 & 3.03 & $+0.01$ & $\Delta\delta,\Delta\omega,\Delta\omega_g$\\
\bottomrule
\end{tabular}
\end{table}

\begin{figure}[t]
\centering
\includegraphics[width=\columnwidth]{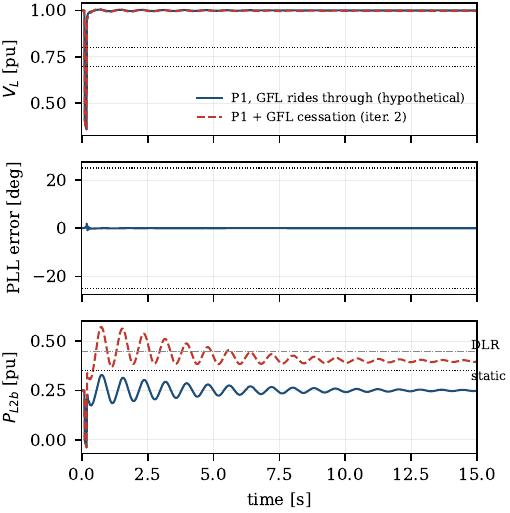}
\caption{P1 transient. Solid: hypothetical case in which the GFL rides through. Dashed:
iteration 2 with legacy GFL momentary cessation. Top: load-bus voltage with 0.8/0.7~p.u. lines.
Middle: PLL angle error with the $\pm25^\circ$ threshold. Bottom: L2b flow against static and
short-duration DLR ratings.}
\label{fig:transient}
\end{figure}

The remaining benchmarks pass, but some only narrowly (Fig.~\ref{fig:transient}, solid lines).
After clearing, the minimum load-bus voltage is 0.981~p.u. The voltage at 20~s is 1.000~p.u.,
and no time is spent below 0.80~p.u. after clearing. The maximum frequency deviation is
0.27~Hz, and the PLL angle error peaks at only $1.9^\circ$, far from $25^\circ$. Using
\eqref{eq:cct}, the SG has $\delta_0=22.0^\circ$, $\delta_{cr}=92.1^\circ$, and
$t_{cr}=0.340$~s (20.4 cycles), giving CCT margins of 15.4 cycles for normal clearing and
5.4 cycles for P5-type backup clearing. Under P5 clearing, the L2b flow peaks at 0.43~p.u. and
the PLL error at $8.0^\circ$. The decisive observation, however, is the \emph{during-fault}
voltage of 0.358~p.u. at the GFL terminal, which is below the legacy $V_{MC}$. Step 7 therefore
infers \emph{GFL momentary cessation without recovery}.

\subsection{Iterations 2--3: propagation to a cascade}
Case S2 removes the GFL dynamics and applies the 0.30~p.u. output loss as a net-load step.
Interestingly, damping improves slightly to 3.03\% because the PLL's contribution to the
weak-corridor coupling disappears. The thermal benchmark now fails instead. L2b peaks at
0.57~p.u. and settles at 0.399~p.u., above its 0.35~p.u. static rating (Fig.~\ref{fig:transient},
dashed). L1 settles at 0.636~p.u. With its thermal overload timer expired, L2b is predicted to
trip (iteration 3). The SG must then supply 1.0~p.u. through L1, which exceeds both its static
(0.75) and DLR (0.90) ratings, so the next actions are L1 tripping and/or UVLS at the load bus.
The sequence
\[
\begin{aligned}
&\text{P1 (L2a)}\to\text{GFL cessation}\\
&\quad\to\text{L2b thermal trip}\to\text{L1 trip/UVLS}
\end{aligned}
\]
contains three automatic actions after the initiating event and meets the probable-cascade
criterion.

\subsection{Causal discovery results}
\begin{figure*}[t]
\centering
\includegraphics[width=0.82\textwidth]{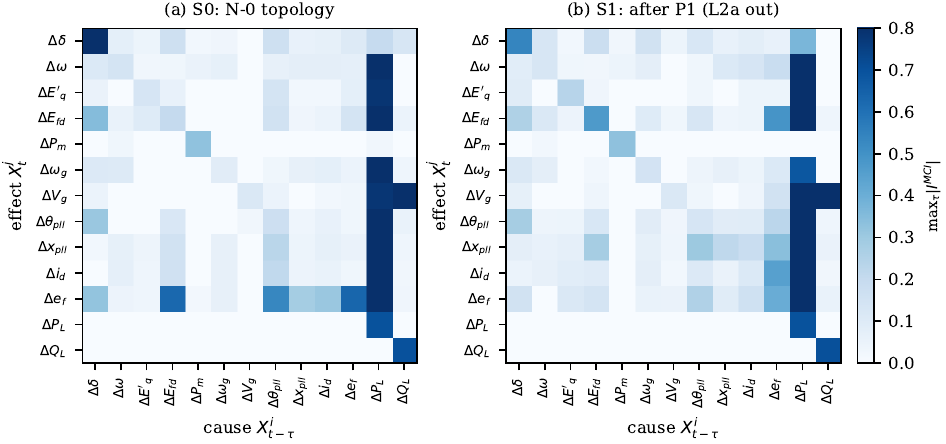}
\caption{PCMCI causal graphs as heat maps of $\max_\tau|I^{MCI}_{i\to j}(\tau)|$ for significant
links (FDR $q\le0.01$). Columns are causes and rows are effects. (a) N-0 topology. (b) After the
P1 outage of L2a.}
\label{fig:mci}
\end{figure*}

\textit{Graph recovery.} Fig.~\ref{fig:mci} shows the MCI strengths for S0 and S1. Against
the ground truth implied by the standardized one-step coefficients of
$[\mathbf{A}_d\ \mathbf{B}_d]$ ($|\cdot|>0.02$), PCMCI recovers adjacencies at any lag with
recall and precision of 0.88/0.62 for S0 and 0.89/0.62 for S1. At lag 1 only, the figures are
0.81/0.66 and 0.82/0.62. Links at lags 2--3 (77 and 63 in S0; 83 and 71 in S1) arise because
the 1\% measurement noise makes the observed process only approximately Markov. This is
expected and supports a small $\tau_{\max}>1$ rather than $\tau_{\max}=1$.

\textit{Topology-dependent causal pathways.} In S0, the strongest parents of
$\Delta\omega_t$ are $\Delta P_{L,t-1}$ ($I^{MCI}=-0.84$), its own lags, and $\Delta\delta_{t-1}$.
In S1, new links appear from the GFL states to the SG speed:
$\Delta e_{f,t-1}\!\to\!\Delta\omega_t$ ($+0.19$) and $\Delta i_{d,t-1}\!\to\!\Delta\omega_t$ ($-0.13$).
In addition, $\Delta e_f$ becomes the dominant state-parent of $\Delta i_d$ ($-0.45$, compared with
$\Delta\theta_{pll}\to\Delta i_d$ of $+0.21$ in S0). The graph therefore shows directly that the
outage couples the GFL synchronization loop into the electromechanical dynamics. Participation
factors alone do not reveal this coupling: in S1, the PLL states have $p_{ik^\star}<0.02$.

\textit{Confounder screening.} For the critical S1 mode, the input participation is
$\eta_{P_L}\approx1.00$ and $\eta_{Q_L}\approx0$. PCMCI finds $\Delta P_L$ links into all three
dominant states, $\Delta\delta$, $\Delta\omega$, and $\Delta\omega_g$ ($I^{MCI}=-0.37,-0.85,-0.68$),
so $\Delta P_L$ is flagged as a confounder. To demonstrate the consequence, we re-ran PCMCI
on S1 with $\Delta P_L$ and $\Delta Q_L$ removed. The state-to-state graph changes from 218 to
104 links, and 26 of these links are not present in the full graph and are therefore spurious.
Omitting the confounder both hides true pathways and invents false ones. This is the failure
mode that assumption (1) and the screen of Section~\ref{sec:rcci} guard against.

\begin{figure}[t]
\centering
\includegraphics[width=\columnwidth]{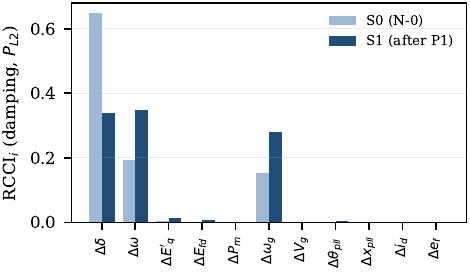}
\caption{Resiliency Causal Contribution Index for the critical mode, using the L2b flow as
the damping benchmark quantity, before (S0) and after (S1) the P1 outage.}
\label{fig:rcci}
\end{figure}

\textit{RCCI.} Fig.~\ref{fig:rcci} shows $\RCCI^{(\zeta)}$ with the L2b flow as the damping
benchmark quantity. In S0, causal responsibility for the critical mode is concentrated in the
SG angle ($\Delta\delta$: 0.65), followed by $\Delta\omega$ (0.19) and $\Delta\omega_g$ (0.15).
After P1, it redistributes to $\Delta\delta$ 0.34, $\Delta\omega$ 0.35, and $\Delta\omega_g$ 0.28,
with excitation ($\Delta E'_q+\Delta E_{fd}$) at 0.02. The share attributable to speed and GFM
frequency, the \emph{damping} channels, rises from 34\% to 63\%. This points to damping-torque
remedies (Table~\ref{tab:map}) as at least as important as synchronizing-strength remedies.
The voltage-benchmark RCCI (load-bus voltage) has nearly the same allocation (0.35/0.35/0.27)
because both benchmarks are carried by the same mode.

\subsection{Expansion alternatives}
Guided by the RCCI and Table~\ref{tab:map}, six alternatives were evaluated at the first
contingency (Table~\ref{tab:alts}). Each piece of equipment is present before and after the
fault. Two outcomes are possible for the GFL. If the during-fault voltage stays above $V_{MC}$,
the GFL rides through and the corridor flow returns to about 0.25~p.u. Otherwise, the
0.30~p.u. loss must be absorbed.

\begin{table}[t]
\centering
\caption{Expansion alternatives evaluated after the P1 contingency.}
\label{tab:alts}
\footnotesize
\setlength{\tabcolsep}{1.8pt}
\begin{tabular}{@{}lcccccc@{}}
\toprule
Alternative & $\zeta_{\min}$ & $V_{\mathrm{fault}}$ & GFL & $P_{L2b}^{ss}$ & Rating & Cascade\\
 & [\%] & [p.u.] & rides? & [p.u.] & [p.u.] & interrupted?\\
\midrule
None (S1)                  & 2.92 & 0.358 & no  & 0.399 & 0.35 & no\\
DLR on L2b                 & 2.92 & 0.358 & no  & 0.399 & 0.45 & no ($\zeta$)\\
STATCOM + POD   & 5.74 & 0.594 & yes & 0.250 & 0.35 & \textbf{yes}\\
Series comp.\ (40\%) & 4.08 & 0.425 & no  & 0.392 & 0.35 & no (thermal)\\
New third L2 circuit       & 4.53 & 0.451 & no  & 0.391 & 0.70 & \textbf{yes}\\
VSC-HVDC + mod.\ & 7.59 & 0.503 & yes & 0.250 & 0.50 & \textbf{yes}\\
\bottomrule
\end{tabular}
\end{table}

The results coincide with the causal attribution. A DLR acts only on the post-action thermal
quantity: it prevents the L2b trip but leaves $\zeta=2.92\%$, so the damping violation
remains. Series compensation strengthens the corridor angle channel ($\Delta\delta$) and restores
$\zeta=4.08\%$, but it increases the voltage during-fault only to 0.425~p.u. The GFL still ceases,
and the thermal problem reappears. A new third circuit fixes both damping and thermal capacity,
although the GFL still stops. The STATCOM with power-oscillation damping and the VSC-HVDC
terminal act on the damping channels ($\Delta\omega,\Delta\omega_g$), which carry 63\% of the
post-contingency RCCI. They also stiffen the load-bus voltage enough for the legacy GFL to ride
through, which removes the first protective action and stops the cascade at its origin.

% ---------------------------------------------------------------------------
\section{Discussion}
\label{sec:disc}

\textit{Scaling to RTO models:} RTO dynamic models contain $10^4$--$10^5$ states. PCMCI
complexity grows roughly as $O(N^2\tau_{\max})$ conditional-independence tests, each costing
$O(T|\mathcal{Z}|^2)$. Only the states with
significant $p_{ik}$ in modes whose damping margin is small, together with the inputs flagged
by $\eta_{jk}$, enter PCMCI, which typically leaves tens to a few hundred variables per study
area.
Optimized PCMCI implementations such as the open-source Tigramite package are available.

\textit{Linearity and fault realism:} Faults, current limiting, and protective actions are
nonlinear. In the proposed method, the small-signal model provides the causal and modal
structure, and the recorder-calibrated fault-equivalent inputs provide benchmark screening.
Every alternative recommended in step 10 should be confirmed by positive-sequence or EMT
simulation before it enters a corrective action plan. The iterative relinearization around
post-action equilibria is a practical compromise between these two needs.

\textit{What the causal claims imply:} PCMCI on model-generated data recovers the causal
structure \emph{of the validated model}. The method therefore strengthens, and does not
replace, model validation against PMU data. The same pipeline can be run on field PMU data,
and disagreement between the field graph and the synthetic graph is itself a useful
diagnostic of model error.

\textit{Limitations of the case study:} The test system is small and the parameters were chosen
to exhibit minimal intrinsic damping. The quantitative values, such as the 3\% crossing and the cessation threshold,
illustrate the procedure and are not general findings. Applying the method to a validated RTO
case with real PMU and DFR archives is the subject of future research.

% ---------------------------------------------------------------------------
\section{Conclusion}
\label{sec:concl}

This paper proposes a resiliency-prioritized transmission expansion planning method that
combines PCMCI causal discovery with eigenvalue participation analysis of PMU-validated
small-signal models of hybrid SG/GFM/GFL systems. Generating synthetic data from validated,
ZOH-discretized state-space models driven by measured load variability satisfies the PCMCI
assumptions by construction. These assumptions are causal sufficiency, realistic forcing,
no contemporaneous causation, a known maximum lag, and a known functional form. The proposed
RCCI quantifies how much each state contributes to each TPL benchmark violation. Input
participation identifies confounders. Iterating protective-action inference with model
rebuilding exposes cascading sequences that a single-step contingency screen misses. The case
study shows that causal attribution separates remedies that only treat a symptom, such as a DLR
alone, from those that interrupt the cascade at its first step, such as damping-capable
FACTS/HVDC and added corridor capacity. Causal inference and PCMCI are therefore valuable tools
for quantifying the causal effects that drive cascading outages. Participation factors help
planners to select the types of equipment that will provide the greatest increase in resiliency.

% ---------------------------------------------------------------------------
\section*{Code Availability}
The Python scripts that reproduce every number and figure in Section~\ref{sec:case} (model,
PCMCI implementation, experiments) accompany this preprint.

% ---------------------------------------------------------------------------
\balance

\end{document}